\documentclass[prl,aps,twocolumn,showpacs,floatfix]{revtex4}

\usepackage{epsfig}
\usepackage{graphicx}
\usepackage{dcolumn}
\usepackage{bm}
\usepackage{amssymb,amsmath}

\begin{document}

\title{Dynamical Mean Field Theory with the Density Matrix Renormalization Group} 

\author{Daniel J. Garc\'{\i}a\footnote{Present address: DFMC, Unicamp, Campinas, S\~ao Paulo, Brasil.}, Karen Hallberg}
\affiliation{Instituto Balseiro and Centro At\'omico Bariloche, 8400 Bariloche, 
Argentina}
\author{Marcelo J. Rozenberg}
\affiliation{CPhT, Ecole Polytechnique, 91128 Palaiseau Cedex, France. \\ Departamento de F\'{\i}sica, FCEN, Universidad de Buenos Aires, 
Ciudad Universitaria Pabell\'on 1, Buenos Aires (1428), Argentina.}

\date{\today}

\begin{abstract}
A new numerical method for the solution of the Dynamical Mean Field 
Theory's self-consistent equations is introduced.
The method uses the Density Matrix Renormalization Group technique to solve the associated impurity problem.
The new algorithm makes no a priori approximations and is only limited
by the number of sites that can be considered.
We obtain accurate estimates of the critical values of the
metal-insulator transitions and provide
evidence of substructure in the Hubbard bands of the correlated metal.
With this algorithm, more complex models having a larger number of degrees of 
freedom can be considered and finite-size effects can be minimized.
\end{abstract}

\maketitle

Great theoretical progress in our understanding of the physics
of strongly correlated electron systems has been possible since the
introduction of the Dynamical Mean Field Theory (DMFT) just over ten
years now \cite{gk,review}.
This approach is based on the natural extension of the familiar
classical mean-field theory of statistical mechanics
to the treatment of models of strongly interacting electrons on a 
lattice.
The DMFT solution of the model is exact in the limit of large lattice
dimensionality or large connectivity \cite{mv,review}.
Since its introduction, DMFT has been widely adopted and was
used for the investigation of a large variety of model Hamiltonians
relevant for problems as diverse as colossal magneto-resistance, heavy 
fermions, metal-insulator transitions, etc. \cite{review}.
A great deal of interest is currently centered around the
ongoing efforts to incorporate DMFT as the local correlation 
physics ``engine'' for first-principle calculations of 
realistic compounds \cite{pu,pt}. 
At the heart of the DMFT method is the solution of an associated
quantum impurity model where the environment of the impurity has
to be determined self-consistently.
Therefore the ability to obtain reliable DMFT solutions of lattice model
Hamiltonians relies directly on the ability to solve quantum impurity
models.
Since solutions of general impurity models are usually not analytically
tractable, one has to resort to numerical algorithms or approximate
methods.
Among the a priori exact numerical algorithms available we count the Hirsch-Fye Quantum Monte Carlo \cite{hf} method and Wilson's Numerical Renormalization Group (NRG) \cite{wrg}.
The former is a finite-temperature method that is formulated in
imaginary time and has been applied to a large variety of impurity
models including the multi-orbital case that corresponds to correlated
multi-band lattice Hamiltonians \cite{mr}.
While this method is very stable and accurate having allowed
for extremely detailed investigations of fundamental problems such as
the metal-insulator transition in the Hubbard model \cite{rchk,review},
its main drawback is that the access to real frequency quantities such
as spectral functions requires to recourse to less controlled
techniques for the analytic continuation of the Green functions.
The second numerical  method is based on 
Wilson's renormalization group \cite{bulla,vollhardt}.
This method can be formulated both at $T$=0 and finite (small) $T$ and 
provides extremely accurate results at very small frequencies, thus it is very adequate for the investigation of correlated metallic phases with 
heavy effective-mass quasi-particles.
The cost to pay is, however, that the description of the high energy
features involve some degree of approximation and cannot
be so accurately obtained \cite{raas}.

The goal of the present work is to introduce a new algorithm for the
solution of the DMFT self-consistent equations, that makes use
of another powerful numerical methodology for the solution of many-body
Hamiltonians: the Density Matrix Renormalization Group (DMRG)\cite{dmrg}.
This method, like the NRG, has the appealing feature
of making no a priori approximations and the possibility
of a systematic improvement of the quality of the solutions.
However, unlike NRG, it is not formulated as a low-frequency asymptotic 
method and thus provides equally reliable solutions for both gapless and
gapfull phases.
More significantly, it provides accurate estimates for the
distributions of spectral intensities of high frequency features
such as the Hubbard bands, that are of main relevance for analysis
of x-ray photoemission and optical conductivity experiments.

We shall illustrate the new formulation with the solution of the, by now classic,
Mott transition in the Hubbard model. We shall show that accurate estimates of the
critical values of the interaction for the metal-to-insulator and for the
insulator-to-metal transitions at $T$=0 can be obtained, and, more interestingly,
we shall provide evidence of substructure in the Hubbard bands in the correlated
metallic phase. 

The Hamiltonian of the Hubbard model is defined by
\begin{equation}
H=t\sum_{\langle i,j\rangle, \sigma} c_{i,\sigma}^{\dagger} c_{j,\sigma}
+ U\sum_i n_{i,\downarrow} n_{i,\uparrow}
\label{hubbard}
\end{equation}
The treatment of this model Hamiltonian with DMFT leads to a
mapping of the original lattice model onto an associated
quantum impurity problem in a self-consistent bath. In the
particular case of the Hubbard model, the associated impurity
problem is the single impurity Anderson model (SIAM), where the
hybridization function $\Delta(\omega)$, which 
in the usual SIAM is a flat density of states of the conduction 
electrons,
is now to be determined self-consistently.
More precisely, for the Hamiltonian (\ref{hubbard})
defined on a Bethe lattice
of coordination $d$, one takes the
limit of large $d$ and exactly maps the model
onto a SIAM impurity problem with
the requirement that $\Delta(\omega) = t^2 G(\omega)$, where
$G(\omega)$ is the impurity Green's function.
At the self-consistent point $G(\omega)$ coincides with the {\em local}
Green's function of the original lattice model \cite{review}.
We take the half-bandwidth of the non interacting
model as unit of energy, thus $t=1/2$.

A central quantity in this algorithm is  the non
interacting Green's function of the impurity problem,
\cal{ G}$_0(\omega) = 1/ (\omega +\mu - \Delta(\omega)) =
1/ (\omega +\mu - t^2 G(\omega))$.
Thus, to
implement the new algorithm we shall consider \cite{qimiao, mpl} a
general representation of the hybridization function in terms of continued
fractions that define a {\em parametrization} of $\Delta(\omega)$ in terms of a
set of real and positive coefficients.
Since it is essentially a Green's function, $\Delta(z)$ can be
decomposed into ``particle'' and  ``hole'' contributions as
$\Delta (z)=\Delta^>(z)+\Delta^<(z)$ with
$\Delta^>(z)=t^2\langle gs|c \frac{1}{z-(H-E_0)} c^{\dagger} |gs
\rangle$ and
$\Delta^<(z)=t^2\langle gs|c^{\dagger}\frac{1}{z+(H-E_0)} c |gs
\rangle$
for a given Hamiltonian, H with ground-state energy $E_0$.
By standard Lanczos technique, H can be in principle
tri-diagonalized and
the functions $\Delta^>(z)$ and $\Delta^<(z)$ can be expressed in
terms of respective continued fractions\cite{Karen}.
As first implemented in Ref.\cite{qimiao,mpl}, each continued fraction
can be represented by a chain of auxiliary atomic sites whose energies
and hopping amplitudes are given by the continued fraction diagonal and 
off-diagonal coefficients respectively.

From the self-consistency condition, the two chains representing
the hybridization, are ``attached'' to the right and left of
an atomic site to obtain a new SIAM Hamiltonian, H.
In fact \cal{G}$_0(z)$ constitutes  the local Green's function of the site 
plus chain system. 

The algorithm in Ref.\cite{qimiao,mpl}, basically consists in
switching on the local Coulomb interaction at the impurity site
of the SIAM Hamiltonian and use the Lanczos technique to re-obtain
$\Delta(z)$, iterating the procedure until the set of
continued fractions coefficients converges.

A great limitation of this procedure is that the number of
auxiliary atomic sites that needs to be used in the hybridization
chains is too large for standard exact diagonalization schemes.
Therefore, the chains have to be cut at very short lengths and as a consequence the method becomes approximate with sizable finite-size effects.
In the present work, we shall make a fundamental improvement on this
scheme, namely diagonalizing the Hamiltonian using DMRG, which in
principle allows to handle chains of arbitrary length \cite{gebhard}.

The SIAM Hamiltonian therefore reads
\begin{eqnarray}
H &=& \sum_{\substack{\sigma,\alpha=-N_C\\ \alpha\ne0}}^{N_C}
 a_{\alpha} c_{\alpha\sigma}^{\dagger} c_{\alpha\sigma}
+\sum_{\substack{\sigma,\alpha=-(N_C-1)\\ \alpha\ne0,-1}}^{N_{C}-1} b_\alpha (c_{\alpha\sigma}^{\dagger} c_{ \alpha+1\sigma}+h.c. ) \nonumber\\
 &&+\sum_{\sigma,\alpha=\pm1} b_0 (c_{\sigma}^{\dagger} c_{\alpha\sigma} + h.c.) + U (n_{ \uparrow}-\frac{1}{2}) (n_{ \downarrow}-\frac{1}{2})
\end{eqnarray}
with $c_{\sigma}$ being the destruction operator at the impurity site,
and $c_{\alpha\sigma}$ being the destruction
operator at the $\alpha$ site of the hybridization chain
of $2 N_C$ sites.
The set of parameters $\{ a_{\alpha}, b_\alpha \}$ are directly
obtained
from the coefficients of the continued fraction representations of
$\Delta(z)$ by the procedure just described.

An important point to make is that while the hopping to an impurity
connected to a chain bears a resemblance to the NRG method, the present
formulation is different in a crucial aspect: unlike NRG, it is not
constructed as an asymptotically exact representation for low
frequencies, but rather treats all energy scales on equal footing.
By paying the price of giving up the excellent resolution at low
frequency of NRG, we shall gain in exchange a controlled and systematic
algorithm operating at {\em all} energy scales.
Of course, as in Wilson's NRG the energy resolution
depends on the length of the auxiliary chain, therefore
considering longer chains while keeping the numerical accuracy
in the computation is the central limitation of the current scheme.
In practice, systems of up to $45$ ($N_C=22$) sites have been considered. 

\begin{figure}[ht]
\includegraphics[width=8cm,clip]{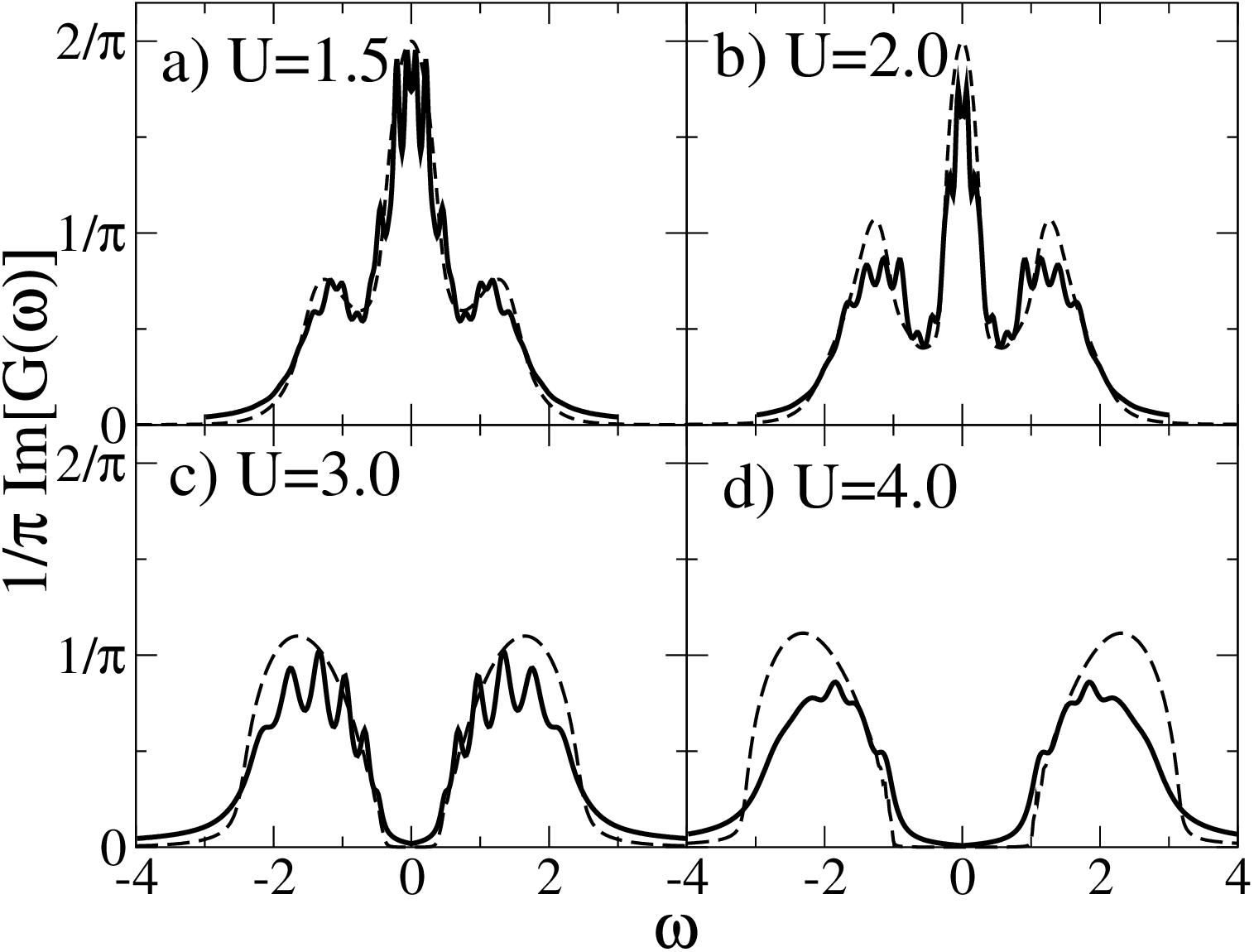}
\caption{Density of states ($\frac{1}{\pi}$Im$G(\omega)$) for 
the half-filled Hubbard Model \cite{pinning}. 
We also show the IPT results (dashed lines). 
\label{fig1}} 
\end{figure}
\begin{figure}[ht]
\includegraphics[width=7cm,clip]{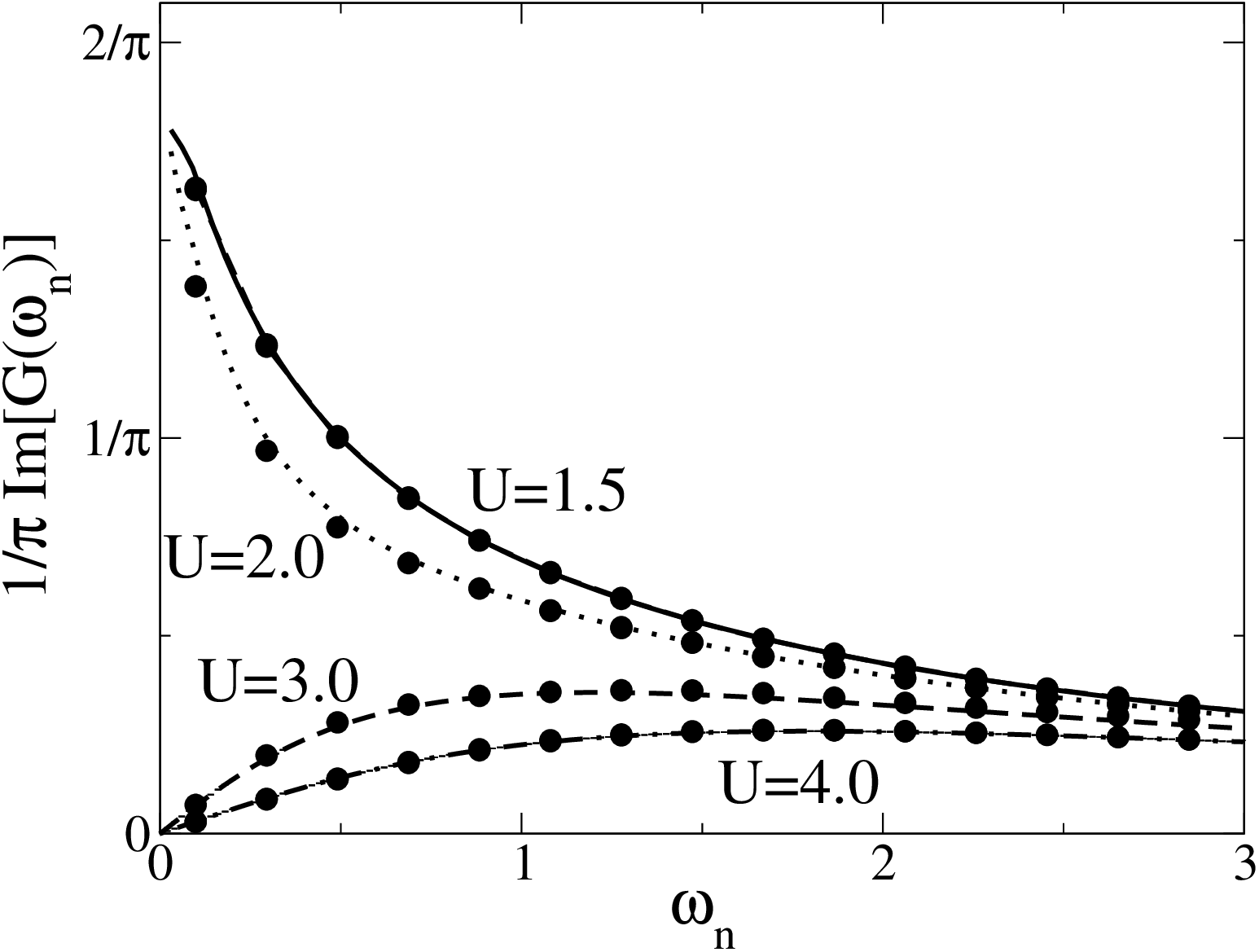}
\caption{Imaginary part of the Green functions at imaginary (Matsubara)
frequencies (solid lines).
We also show  Quantum Monte Carlo results (circles) at low temperature $T=1/32$. 
\label{fig2}}
\end{figure}
\begin{figure}[ht]
\includegraphics[width=8cm,clip]{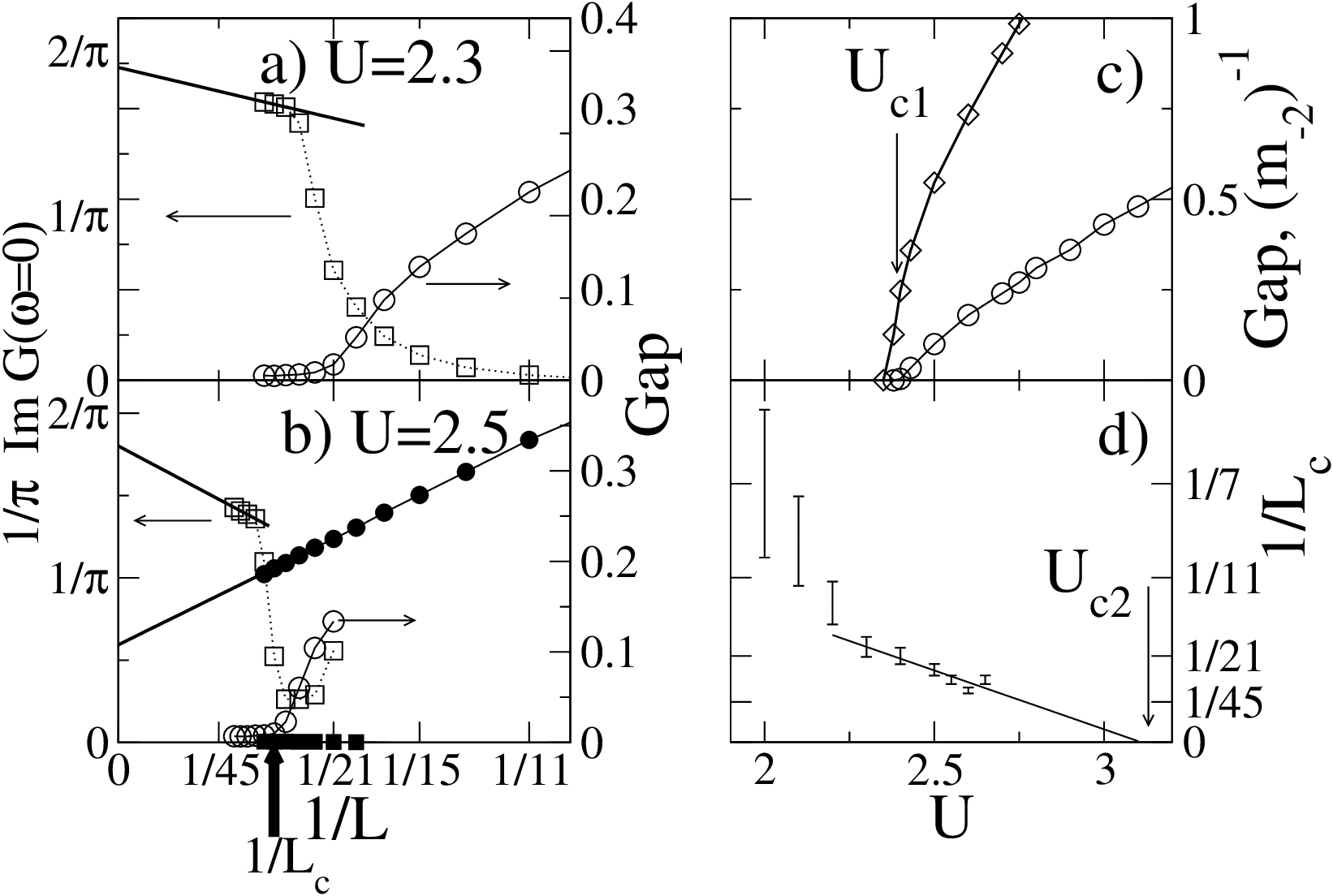}
\caption{Density of states at the Fermi energy, $\frac{1}{\pi}$Im$G(\omega=0)$, (squares) and gap (circles) for a) $U=2.3$ and b) $U=2.5$.
Open symbols correspond to metallic solutions and closed symbols in (b) to the insulating phase.
The thick arrow shows the critical length ($L_c$) above which a metallic solution is obtained.
c) Gap (circles) and inverse of the second moment of the DOS (diamond) as a function of interaction $U$. 
The values correspond to extrapolations to infinite-size chains.  
d) Critical length $L_c$ as a function of interaction $U$. \label{fig3}}
\end{figure}

We now turn to the results of the implementation of this algorithm   for
the solution of the DMFT equations of the Hubbard model.
In Fig.\ref{fig1} we show the DMFT+DMRG results (solid lines) for the
density of states (DOS) for several values of increasing interaction U.
The results are compared to the Iterated Perturbation Theory (IPT)
results (dashed lines) \cite{gk,zrg} which is a useful analytic
approximate method that can be solved on the real frequency axis at
$T=0$.
In Fig.\ref{fig2} we compare the DMFT+DMRG results (solid lines) for the
imaginary part of the Green's function on the imaginary frequency axis
with the corresponding ones obtained by quantum Monte Carlo solution of
the DMFT equations at a low temperature (circles).
We see that the overall agreement on the real axis is very satisfactory
and on the Matsubara frequency axis it is excellent.
At low values of the interaction $U$ we find a metallic state characterized by a
narrow quasi-particle feature at low frequencies. Increasing $U$ this peak gets 
narrower by transferring spectral
weight to features at higher energies of order $U$, the upper and lower
Hubbard bands.
At large values of the interaction the system evolves toward an
insulating state with a gap of order $U$ (Fig. \ref{fig3}(c)).

A very important feature of the metal-insulator transition 
in the paramagnetic state of the Hubbard model at half-filling
\cite{review} is that there are two distinct critical values of the
interaction associated with the transition: $U_{c1}$ and $U_{c2}$.
The former signals the insulator to metal transition obtained upon
lowering the interaction, while the latter
corresponds to the metal to insulator transition obtained when the Fermi
liquid is destroyed by increasing the interaction strength.
We obtained estimates of these two values that are consistent with
those from NRG calculations \cite{bulla}.
We find $U_{c1}=2.39 \pm 0.02$ and $U_{c2}=3.0 \pm 0.2$.
Due to the nature of this algorithm and the arguments presented
before, we expect that our determination of $U_{c1}$ should be
more accurate than NRG (and all other previously used $T=0$ methods).
Our criterion for the investigation of metallic versus insulating states
was based on the behavior of the spectral weight at zero frequency and
the size of the gap in the DOS (given by the energy of the first pole).
It was a remarkable finding that these quantities showed an unexpected
dramatic dependence with the length of the hybridization chain and the
proximity to the critical value of the interaction.
This dramatic effect is demonstrated in Fig.\ref{fig3}, where we plot
the results as a function of the inverse of the chain length at $U=2.3$ (a) and 
$U=2.5$ (b).
We find that as we increase the value of $U$ from the weak coupling
(metallic) side, chains longer than a $U$-dependent value $L_c$ were 
required to converge to metallic solutions (Fig.\ref{fig3} d)).
The value of $L_c$ was in fact found to {\em diverge} at a finite
interaction strength which we identified with our estimate of
$U_{c2}$. 
As long as the length $L=N_c < L_c$ the solution 
looks as that of an insulating state with vanishing density
of states at $\omega=0$, while a rapid crossover to a metallic solution
is seen as $L$ goes beyond the threshold value $L_c$.
Interestingly, 
this resembles the  behaviour of the Kondo effect in finite 
systems \cite{cornaglia}, where the Kondo effect is strongly 
suppressed if the Kondo screening cloud is larger than the system size. 
This similarity also shows up in the increase of $L_c$ with $U$, since 
in the Kondo model the correlation length increases with the interaction.

On the other hand, for the insulator to metal transition, the value of 
$U_{c1}$ can be determined by the closure of the gap in the DOS or using 
the inverse of the second moment of the DOS\cite{rozMom}, Fig. \ref{fig3} c).  
Any of these two quantities are non zero in the insulating state 
and vanish at $U_{c1}$.

\begin{figure}[ht]
\includegraphics[width=8cm,clip]{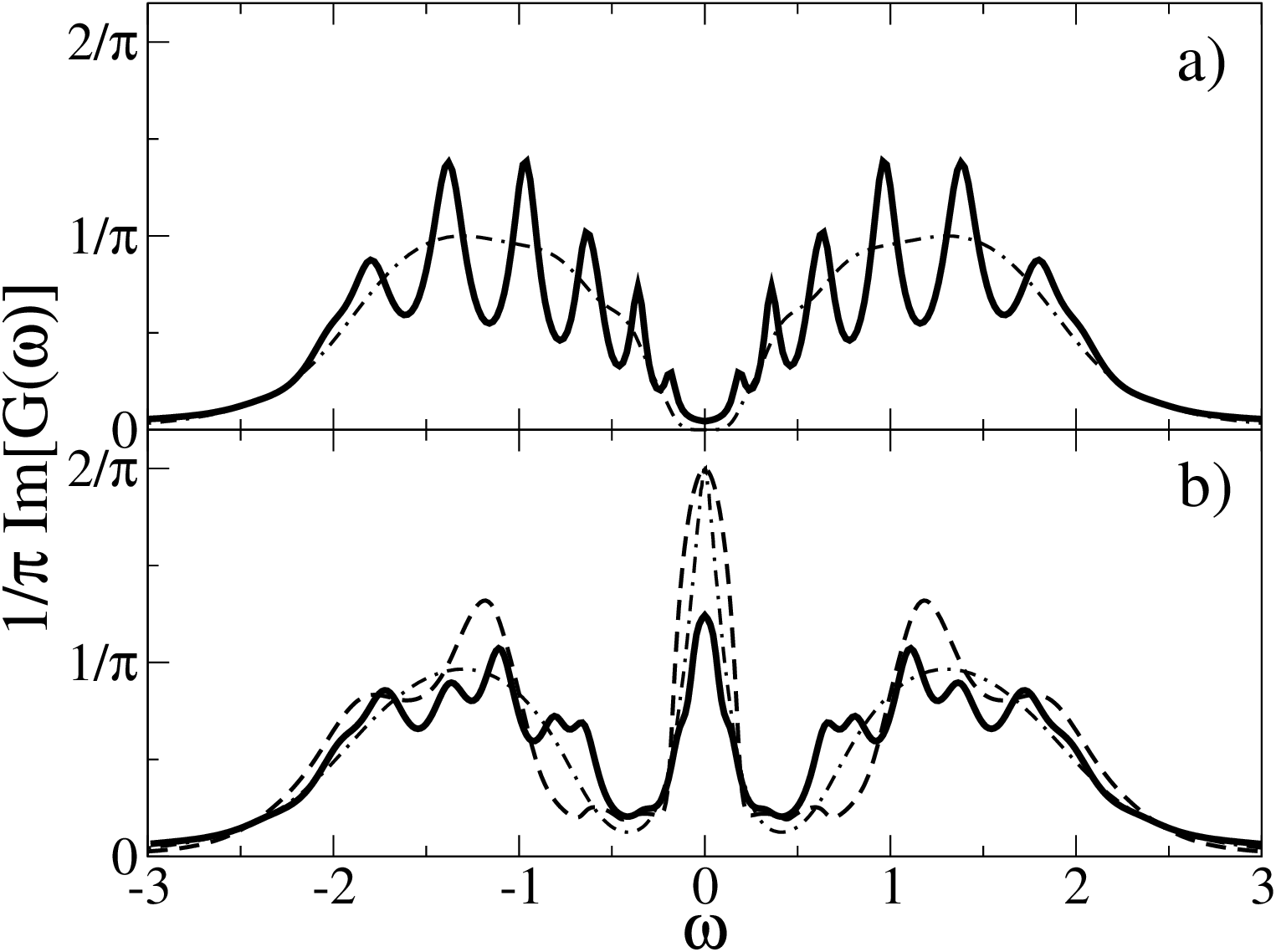}
\caption{Density of states for U=2.5 (solid line).
For comparison NRG (dot-dashed line) and IPT (dashed line) results 
are also shown. a) Insulating solution. b) Metallic solution.\label{fig4}}
\end{figure}

Another interesting result that has been a matter of debate, and might have
implications for the analysis of x-ray photoemission spectra, is the question of
the existence on substructure in the Hubbard bands of the correlated metallic
state.  
The substructure was first identified within IPT calculations \cite{zrg},
but the exact numerical methods did not have the required accuracy to decide
whether the substructure was an artifact borne out of IPT or a real feature of
the model's solution.  In fact, an appealing physical interpretation of the
substructure can be readily made: in a rigid band picture, one can think of the
action of the local interaction $U$ to ``split'' and replicate the hybridization
function density of states $\Delta(\omega)$ at frequencies $\pm U/2$. This in
fact gives a simple qualitative understanding of the emergence of Hubbard bands
in the insulator when $U$ is large. The semicircular $\Delta(\omega)$ is
duplicated into the two (approximate) semicircles at $\pm U/2$ of the local DOS
(the imaginary part of $G(\omega)$) characteristic of the Mott-Hubbard insulator.  
Then one realizes that the self-consistency requires $\Delta(\omega)$ to coincide
with $t^2G(\omega)$, thus, if Im$[G(\omega)]$ develops (multi-peak) structure it
implies that $\Delta(\omega)$ has a similar structure. Then, by the simple-minded
argument in which the action of $U$ is to ``split and replicate'', we get
(multi-peak) structures for the Hubbard bands at $\pm U/2$ in the DOS, in a
self-consistent manner.  In Fig.\ref{fig4} we show the comparison of the DOS of
the model for both the correlated metallic and insulating solutions at the same
value of the interaction $U$=2.5 (i.e., between $U_{c1}$ and $U_{c2}$). 
The results clearly show that the smooth envelope of
the pole structure (due to a finite $N_c$) forming the lower and upper 
Hubbard bands in the insulator,
acquires more structure in the metallic state.
For further comparison, we also plot the corresponding results
from NRG\cite{bullapriv} and IPT.
The former fails to capture any qualitative difference between the shape of the 
metallic and insulating Hubbard
bands, however, IPT shows a very smooth shape for the insulator (cf
Fig.\ref{fig1}c-d) and a multi-peak structure in the metallic Hubbard bands. 
The observed structure of IPT is in qualitative agreement with the spectral 
 distribution of weight born-out from the DMRG calculation.  To our knowledge this is the first strong evidence of the existence of 
 non-trivial structure in the Hubbard bands within DMFT. 

To conclude, we have presented a new algorithm to solve the
DMFT equations of strongly correlated models exploiting 
the DMRG methodology. 
Large systems can be considered and accurate values of the critical 
interactions are obtained in agreement with NRG predictions allowing 
for a non-trivial test of the accuracy of this method. 
In contrast with NRG, however, this new algorithm deals with all 
energy scales on equal footing which allowed us to find interesting 
substructure in the Hubbard bands of the correlated metallic state. 
This observation might be useful for the interpretation of high 
resolution photoemission spectroscopies. 
In addition this method can handle more general models having 
a larger number of degrees of freedom and can be
generalized to finite temperatures.

We thank R. Bulla for the NRG data and B. Alascio for useful discussions. 
We acknowledge support from CONICET, Fundaci\'on Antorchas (14116-168), 
PICT 03-06343 of ANPCyT.



\begin{thebibliography}{20}
\bibitem{pinning} note that the ``pinning'' condition seems not to
be obeyed due to the finite broadening of the $\delta$-function poles.
\end{thebibliography}
\end{document}